\documentclass[12pt]{article}
\usepackage{amssymb}
\usepackage{amsmath}
\usepackage{amsthm}
\usepackage{epsfig}
\usepackage{graphicx}
\usepackage[numbers,sort&compress]{natbib}
\usepackage{rotating}
\usepackage{url}
\usepackage{xcolor}
\usepackage{float}
\usepackage[utf8]{inputenc}
\usepackage[english]{babel}
\usepackage[paper=a4paper]{geometry}
\usepackage{lscape}
\usepackage{latexsym}
\usepackage{slashed, amsthm}

\title{The Replacement Rule for Nonlinear Shallow Water Waves}
\author{Z. Zong$^{1,2,3}$,  A. Ludu$^{4}$
	\\
	\\ \small{1. School of Shipbuilding Engineering, Dalian University of Technology}
	\\ \small{2. Collaborative Centre of Advanced Ships and Deepwater Engineering}
	\\ \small{3. Liaoning Deepwater Floating Structure Engineering Technology Lab}
	\\ \small{4. Department of Mathematics, Embry-Riddle Aeronautical University} \\ \small{Daytona Beach, FL, USA}
	\\}

\begin{document}
	
	\maketitle 
	{\let\thefootnote\relax\footnotetext{{\em Emails}:
			zongzhi@dlut.edu.cn, ludua@erau.edu}}

\begin{abstract}
\noindent
When a $(1+1)$-dimensional nonlinear PDE in real function $\eta(x,t)$ admits localized traveling solutions we can consider $L$ to be the average width of the envelope, $A$ the average value of the amplitude of the envelope, and $V$ the group velocity of such a solution. The replacement rule (RR or nonlinear dispersion relation) procedure is able to provide a  simple qualitative relation between these three parameters, without actually solve the equation. Examples are provided from KdV, C-H and BBM equations, but the procedure appears to be almost universally valid for such $(1+1)$-dimensional nonlinear PDE and their localized traveling solutions \cite{3}. 
\end{abstract}

\section{Introduction}

Essential in water waves is that waves exist only when the wave parameters (wave frequency, wave number and/or amplitude ) satisfy the dispersion relation \cite{1,2}. Therefore, dispersion relation is the central part in water wave theory. The dispersion relation in linear water wave theory has been well established. The Linear Superposition Principle helps us to decompose a linear wave train into a sum of infinitely many sinusoidal/cosine components. In other word, the solution to the wave evolution equation can be written as the sum of infinitely many sinusoidal and/or cosine functions. This permits us to focus only on one component wave $\eta(x,t)\eta(x-ct)=\mathit{R}Ae^{i(kx-\omega t)}$, where $\eta(x,t)$ is wave elevation, $\mathcal{R}$ represents the real part, $A$ is the amplitude, $k$ and $\omega$ are wave number and frequency respectively. Using it, we obtain a simple replacement relation in the form of
\begin{equation}\label{eq1}
\eta_t \rightarrow -i \omega, \ \ \eta_x\rightarrow ik, \ \ \eta_{xx}\rightarrow -k^2, \eta_{tt} \rightarrow -\omega^2,\dots ,
\end{equation}
which permits replacing derivatives with respect to time and space in the evolution equation by variables, and changing evolution equation into the algebraic dispersion relation. In the case of finite water depth dispersion relation correlates wave frequency with wave number and water depth in the form of $\omega^2=gk \tanh (kh)$. Here $g$ is gravity acceleration, $h$ is water depth and $\tanh(\cdot)$ is tangent hyperbolic function. In the dispersion formula, amplitude is absent, featuring the property of linear waves. It explains some very interesting phenomena, among which are longer waves traveling faster than short waves in deep water and waves traveling parallel to the shore in shallow water, to name a few \cite{2}.

The dispersion relation in nonlinear water wave theory is much more complicated than its linear counterpart. It is characterized by inclusion of wave amplitude, thus being amplitude dispersion or nonlinear dispersion relation (NLDR). 
It seems extremely difficult to write explicitly the relation for nonlinear waves, and approximate methods are applied. Ludu \& Kevrekidis, \cite{3}, provided a simple but promising methodology to find NLDR for a specified nonlinear evolution equation. In their procedures, derivatives with respect to time and space are replaced with wave amplitude multiplied/divided by characteristic length/time. It is capable of yielding scale relation among the wave parameters. Inspired by the method, we continue the efforts to find the approximate algebraic relation between differential operators and wave parameters such that to obtain approximate yet accurate enough NLDR. 

\section{The replacement Rule (RR)}

We want to generalize the replacement procedures in Eq. (\ref{eq1}) to nonlinear waves in an approximate way which also attaches additional physical meaning to some traditional nonlinear wave equations. This is at least theoretically possible because \cite{3} proved that the convolution of wave elevation at present time and at a much later time instant is zero. That is to say, if the considered temporal interval under consideration is long enough, memory effects in a nonlinear wave are negligible. This makes it possible to find the Replacement Rule (RR) for nonlinear wave waves in the approximate meaning.   

A progressive nonlinear wave profile, as a solution of a nonlinear wave equation of the form
$$
\eta_t = \mathcal{O} (\eta,\eta_x ,\eta_{xx},\eta_{xxx},\eta_{xxt}, \dots),
$$
where $\mathcal{O}$ is a nonlinear integro-differential operator, can be regarded as a wave-packet, which can be further expressed by the sum of infinitely many harmonic waves in the form of 
\begin{equation}\label{eq2}
\eta(x,t)=\int_{0}^{\infty} A(k) \exp[i(kx-\omega t)] dk,
\end{equation}
where $\eta$  is surface elevation and $A(k)$ is the amplitude of each component wave,   $k$ is wave number, $\omega$ is wave frequency in radian, which in principle depends on   $k$, and $i=\sqrt{-1}$ is the imaginary unit. Such general, and often non-local nonlinear equations are not necessary Galilean invariant \cite{7}. The harmonic waves combined in Eq. (\ref{eq2}) can be divided further in classes of nonlinear progressive wave packages traveling with various group velocities. We can hypothetically separate these packages in classes of independent and parallel copies of the $x-$axis, and assume that each such package is followed and recorded by a sensor attached to its co-moving reference frame, and moving with a velocity $V(k)$ . Next, we collect all these signals, from all the sensors, shift each of them with a certain delay $\tau(k)$  and then overlap the signals in the same reference frame. Mathematically, this is equivalent to rewrite Eq. (\ref{eq2}) in the form 
\begin{equation}\label{eq3}
\eta_{P}(t)=\int_{0}^{\infty} A(k) \exp [it(k V(k)-\omega)]dk=\int_{0}^{\infty}A(k) \exp [it G(k)]dk,
\end{equation}
where $G(k)=k V(k)-\omega$ and $V(k)=\frac{x}{t}$, where $V(k)$  is the speed of each harmonic wave in the package, and subscript $P$ indicates progressive wave character. The exponential terms containing the delays of various progressive packages $\tau(k)$ are absorbed into a new form for $A(k)$.

Actually, since we are seeking for stationary solutions, and thus we can choose time $t$  to be large enough, we can even neglect the contributions of these delays. In this asymptotic limit for time the phase in the above integral becomes a highly oscillatory function, which will cancel large contribution of the integrand. Only the region of integration where the exponent is  almost flat contribues to the integral. Since the wavelength of all possible water waves are bounded from below by the smallest possible ripples we can consider the function $A(k)$ to be rapidly decreasing when wave number approaches infinity, or even consider it with compact support on the wave number axis. Also, considering $A(k)$ as being zero for negative we can indeed use the stationary-phase approximation formula. So, by using the stationary-phase approximation, \cite{4}, we obtain

\begin{equation}\label{eq4}
\int_{0}^{\infty} A(k) \exp [it (k V(k)-\omega)] dk \simeq A(k_{0})\sqrt{\frac{2 \pi}{t|\omega^{''}(k_0 )|}}\exp \biggl[it (k_0 V(k_0 )-\omega_0 ) \pm \frac{\pi}{4}) \biggr],
\end{equation}
where
\begin{equation}\label{eq5}
\frac{dG}{dk}\biggl|_{k=k_0}=\biggl( V-\frac{d\omega}{dk}+k \frac{dV}{dk} \biggr)\biggl|_{k=k_0 }=0.
\end{equation}
Here $\pm$  in front of $\pi/4$ is the sign of $\omega^{''}(k_0)$. Note that the derivative of frequency with respect to wave number is the group velocity. So the above equation becomes $c_g =V$ at $k=k_0$. This is in fact required by a shallow water wave theory, and thus the theory in this paper is applicable to shallow water waves.  

We may write the terms in front of the exponential function as a single term, and neglect the subscript $0$ with the understanding that wave number is close to the value ensuring the group velocity is equal to wave speed, that is,   
\begin{equation}\label{eq6}
\eta(x,t)\simeq A(k) \exp \biggl[ i(kx-\omega t)\pm \frac{\pi}{4}\biggr] .
\end{equation}
Although amplitude $A$ contains time inside the square root sign, an implicit assumption is that as time is long enough, the wave $\exp(iks-i\omega t)$ oscillates several cycles while time $t$ changes little. To avoid using complicated double-scale analysis to justify the approximation, we may adopt piece-wise linear approximations to interpret Eq. (\ref{eq6}). We specify a time interval $(T_{n}, t_{n+1})$  in such a way that time $t$   does not change significantly, but $\exp(i k x-i \omega t)$  oscillates several cycles. After that we specify a new interval $(t_{n+1},t_{n+2})$  in which the wave $\exp(i k x-i \omega t)$  oscillates several cycles again. We repeat the procedures, being able to obtain a series of linear waves of varying amplitudes valid on a finite interval which approximate a nonlinear wave:
$$
\eta(x,t)\simeq \sum_{n}^{\infty}A(k,t_{n}) \mathbf{1}_{(t_{n},t_{n+1})} \exp \biggl[ i \biggl( k x- \omega t \pm \frac{\pi}{4} \biggr) \biggr] ,
$$
where $\mathbf{1}_{(t_{n},t_{n+1})}$ is the indicator function of the interval $(t_{n},t_{n+1})$. In other words, we us a linear wave train in the form of Eq. (\ref{eq6}) to approximate a nonlinear wave train within a temporal interval and use another linear wave train to approximate the nonlinear wave train on the subsequent temporal interval. In this way, the specific dependence on the asymptotic time becomes just a parametric dependence, and not a dynamical one.  
     
Keeping such understanding of piecewise linear approximation in mind, we obtain the Replacement Rule (RR) for nonlinear shallow water waves in the form of 
\begin{equation}\label{eq7}
\eta \rightarrow A, \ \  \eta_{t} \rightarrow -i \omega A, \ \  \eta_{x} \rightarrow  i k A, \ \  \eta_{xx}  \rightarrow -k^2 A, \ \  \eta_{tt} \rightarrow -\omega^2 A, \dots 
\end{equation}
The RR is exact for linear waves. In the linear case, we do not have amplitude   appearing in the procedure, for it can be removed from both sides of the evolution equation. But in nonlinear evolution equation we include amplitude   in the RR procedure.
 
Another interesting observation can be obtained from Eq. (\ref{eq6}) through slight change
\begin{equation}\label{eq8}
\eta(x,t)\simeq A(k) \exp [i k t (c_{g} - c_{V})].
\end{equation}
In deep water, the two velocities (group and phase) are significantly different, thus the nominal wave frequency $\Omega = k (c_{g}-c_{V})$ is much larger than its counterpart in shallow water (where group and phase velocities are very close). So in deep water it is not easy to form long-period waves while in shallow water we have. So deep water is characterized by short waves while long-waves are observable near the shore.
 
In the following, we apply the above RR to some equations in nonlinear shallow water waves. 

\section{Examples}

\subsection{KdV Equation}

Although KdV equation is derivable from several ways, they are all rooted in the basic procedures: (1) retaining weakly nonlinear terms in the governing equation, (2) revising the linear dispersion relation to include weak dispersion approximation, and (3) substituting the weakly dispersion relation into the governing equation to obtain the required KdV equation. A subtle paradox inherent with the methodology is that the dispersion relation obtained from the final KdV equation is different from the beginning linear dispersion relation because amplitude dispersion is included in KdV. This paradox deteriorates the theoretical consistency of KdV equation.      

To eliminate the above paradox and remove theoretical consistency, we start from Whitham’s methodology to demonstrate the power of the RR method. The Whitham’s strategy, \cite{5,6}, is simply of the form   
\begin{equation}\tag{8a}
\eta_{t}+\int_{-\infty}^{\infty}K(x-\zeta) \eta_{\zeta} (\zeta ,t) d \zeta =0, 
\end{equation}
\begin{equation}\tag{8b}
K(x)=\frac{1}{2 \pi} \int_{-\infty}^{\infty} c(k) \exp [i k x] dk,
\end{equation}
where phase velocity is
\begin{equation}\label{eq9}
\frac{\omega}{k}=c(k)=c_{0} \sqrt{\frac{\tanh (k h)}{k h}}, \ \ c_{0}=\sqrt{g h}.
\end{equation}
Here we omit the nonlinear term in Eqs. (8) compared with Whitham’s original equations. We will demonstrate in the following that including nonlinear terms in Whitham’s equations is not only unnecessary but also arbitrary if we have the RR at hand. 

Unlike the ordinary derivations, we include amplitude dispersion effect at the very beginning in the form of  
\begin{equation}\label{eq10}
c(k) \simeq c_0 \biggl[ 1-\frac{1}{6}(k h)^2 +\alpha A+ \mathcal{O}(k^4 ) \biggr] .
\end{equation}
In order to remove the theoretical inconsistency. Substituting Eq. (\ref{eq10}) in (8b) yields after some simple integration  
\begin{equation}\label{eq11}
K(x)=c_0 \delta (x)+\frac{1}{6}c_0 h^2 \delta^{''}(x)+\alpha c_0 A \delta (x).
\end{equation}
Simple integrating Eq. (8a) leads to the governing equation  
\begin{figure}[H]
\centering
\includegraphics[scale=1.02]{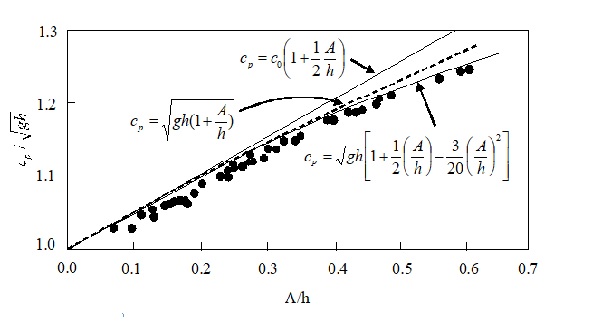} 
\caption{Speed of solitary wave vs. relative amplitude (dots denote experimental results from \cite{15}).}
\label{fig1}
\end{figure}
\begin{equation}\label{eq12}
\eta_t +c_0 \eta_x +\frac{1}{6} c_0 h^2 \eta_{xxx}+c_0 \alpha A \eta_x=0.
\end{equation}
KdV equation follows immediately if we apply the RR Eq. (\ref{eq7}) in the reverse way
\begin{equation}\label{eq13}
\eta_t +c_0 \eta_x +\frac{1}{6} c_0 h^2 \eta_{xxx}+c_0 \alpha \eta \eta_x=0.
\end{equation}
Therefore why is good to do so is to avoid the previously mentioned paradox
and arbitrary introduction of nonlinear terms in the original Whitham's method.

Here in fact we provide a new method to derive governing equation in shallow water. Starting from the dispersion relation Eq. (\ref{eq10}), and substituting it into Whitham’s equations Eq. (\ref{eq8}), we obtain KdV equation based on our Replacement Rule Eq. (\ref{eq7}) in an easy way. This greatly simplifies derivation of nonlinear governing equations of shallow water waves, which used to be very heavy brain workloads for scientists working in this field.   

We can double check the procedure. Indeed, by using the Replacement Rule, we have
\begin{equation}\label{eq14}
-i \omega A + i c_0 A k -i \frac{1}{6}c_0 h^2 A + i k \alpha A^2 =0,
\end{equation}
which turns out to be
$$
\omega= c_0 k -\frac{1}{6}c_0 h^2 k+c_0 \alpha k A,
$$
\begin{equation}\label{eq15}
c_p =\frac{\omega}{k}= c_0 \biggl( 1-\frac{1}{6} c_0 h^2 + \alpha A \biggr),
\end{equation}
exactly the same as our assumption.

\subsection{Camassa-Holm Equation}

The Camassa-Holm (C-H) equation, \cite{12}, describes shallow water waves with discontinuous derivatives. It reads 
\begin{equation}\label{eq16}
u_t +2 \kappa u_x -u_{xxt}+3 u u_{x}=2 u_{x} u_{xx}+u u_{xxx},
\end{equation}
where $\kappa$ is a constant. Then the RR method leads to
\begin{equation}\label{eq17}
-i \omega A+2 i \kappa k-i \omega A k^2+3 i k A^2=-2 i A^2 k^3 - i k^3 A^2.
\end{equation}
Simplifications lead us to
\begin{equation}\label{eq18}
\omega=
\frac{2 \kappa +3 A +3 A k^2}{1+k^2}k
\end{equation}
It reduces exactly to linear dispersion relation as amplitude terms are neglected. The phase velocity is then  
\begin{equation}\label{eq19}
c_p =\frac{2 \kappa +3 A +3 A k^2}{1+k^2}=3A+\frac{2 \kappa}{1+k^2}.
\end{equation}
In the two asymptotic limits, it reduces to
\begin{equation}\tag{20a}
c_p = \rightarrow 2 \kappa +3A,  \ \ \ k \rightarrow 0,
\end{equation}
\begin{equation}\tag{20b}
c_p \rightarrow 3A, \ \ \ k \rightarrow \infty.
\end{equation}
It is surprising that as wave becomes very short, phase velocity is not zero, but tends to be proportional to its amplitude. Larger waves overtake smaller waves, and stronger wins in the final game. The group velocity is  
\begin{equation}\label{eq21}
c_g =\frac{d\omega }{d k}=\frac{1}{(1+k^2)^2}[(2 \kappa +3 A + 9 A k^2)(1+k^2)-2 k^2 (2 \kappa +3 A +3 A k^2)],
\end{equation}
which reduces to the following limiting forms
\begin{equation}\tag{22a}
c_g \rightarrow 2 \kappa +3 A, \ \ \ k\rightarrow 0, 
\end{equation}
\begin{equation}\tag{22b}
c_g \rightarrow 3 A, \ \ \ k \rightarrow \infty.
\end{equation}
Long wave, group and phase velocities are same as expected.  

We now derive C-H equation based on the dispersion relation Eq. (\ref{eq19}). Using Eq. (\ref{eq19}) we have
\begin{equation}\tag{23a}
K(x)=\frac{1}{2 \pi} \int_{-\infty}^{\infty} c(k) \exp [ i k x] dk=3 A \delta(x)+ 2 \kappa \exp (-|x|).
\end{equation}
Then
\begin{equation}\tag{23b}
\int_{-\infty}^{\infty} \exp (-|x-\zeta|)\eta_{\zeta} (\zeta,t) d \zeta =
\int_{-\infty}^{x} \exp (\zeta-x) \eta_{\zeta} (\zeta, t) d\zeta +
\int_{x}^{\infty} \exp (x-\zeta) \eta_{\zeta} (\zeta, t) d\zeta .
\end{equation}
Simple integration by parts yields
$$
\int_{-\infty}^{\infty} \exp (-|x-\zeta|) \eta_{\zeta}(\zeta,t) d\zeta =2 \eta_x +2 \eta_{xxx}+2 \eta_{xxxxx}+\dots =2 \eta_x +2 \frac{\partial^{2}}{\partial x^{2}}(\eta_{x}+\eta_{xxx}+\dots),
$$
\begin{equation}\tag{23c}
\eta_{t}+\int_{-\infty}^{\infty} K(x-\zeta)\eta_{\zeta} (\zeta,t) d\zeta=
\eta_t +3 A \eta_x +2 \eta_x + 4 \kappa \frac{\partial^{2}}{\partial x^{2}}(\eta_x +\eta_{xxx}+\dots )=0.
\end{equation}
Letting $I=\eta_{x}+\eta_{xxx}+\dots$, we may obtain recurrence relation
\begin{equation}\tag{24a}
I=\eta_x + \frac{\partial^{2} I}{\partial x^{2}}.
\end{equation}
From Eq. (24a) we have
\begin{equation}\tag{24b}
4 \kappa \frac{\partial^{2} I}{\partial x^{2}}=-(\eta_t +3 A \eta_x +2 \eta_x),
\end{equation}
or
\begin{equation}\tag{24c}
\eta_t +3 \eta \eta_x +2 \eta_x +4 \kappa \frac{\partial^{2}}{\partial x^{2}} \eta_x +4 \kappa \frac{\partial^{2}}{\partial x^{2}} \biggl( \eta_{t} +\frac{\partial^{2} I}{\partial x^{2}} \biggr)=0,
\end{equation}
which further reduces to
\begin{equation}\label{eq25}
\eta_t +3 \eta \eta_x +2 \eta_x +4 \kappa \eta_{xxx} - \frac{\partial^{2}}{\partial x^{2}} ( \eta_{t} +3 \eta \eta_x + 2 \eta_x)=0.
\end{equation}
The final form is
\begin{equation}\tag{26}
\eta_t +3 \eta \eta_x +(4 \kappa -2) \eta_{xxx}+2 \eta_x-\eta_{xxt}=3 \eta \eta_{xxx}+9 \eta_x \eta_{xx}. 
\end{equation}
If we set $\kappa = 1/2$, we then obtain C-H equation again. We show again here how to use the replacement rule to obtain nonlinear evolution equations for shallow water waves in an concise way.

\subsection{Benjamin-Bona-Mahony (BBM) Equation}

This equation, \cite{13}, was studied in Benjamin, Bona, and Mahony (1972) as an improvement of the Korteweg–de Vries equation (KdV equation) for modeling long surface gravity waves of small amplitude–propagating uni-directionally in  $1+1$ dimensions.  They show the stability and uniqueness of solutions to the BBM equation. This contrasts with the KdV equation, which is unstable in its high wave number components.
 
Further, while the KdV equation has an infinite number of integrals of motion, the BBM equation only has three. Before, in 1966, this equation was introduced by Peregrine, in the study of undular bores.

BBM equations follows if the terms on the right hand side of CH equation are neglected, that is, 
\begin{equation}\tag{27}
\eta_t +\eta_x +\eta \eta_x -\eta_{xxt}=0.
\end{equation}
Using the RR we have
\begin{equation}\tag{28}
-i A \omega + i A k + i k A^2 -i k^2 \omega A =0,
\end{equation}
or dispersion relation reads
\begin{equation}\tag{29}
\omega=\frac{k (1+A)}{1+k^2}.
\end{equation}
Simple operations lead to
\begin{equation}\tag{30}
c_p =\frac{\omega}{k}=\frac{1+A}{1+k^2}, \ \ \ c_g = \frac{d \omega}{dk}=\frac{(1-k^2)(1+A)}{(1+k^2)^2},
\end{equation}
\begin{equation}\tag{30a}
c_p \rightarrow 
\begin{cases}
1+A, & k \rightarrow 0 \\
0, & k \rightarrow \infty
\end{cases}
\end{equation}
\begin{equation}\tag{30b}
c_g \rightarrow 
\begin{cases}
1+A, & k \rightarrow 0 \\
0, & k \rightarrow \infty
\end{cases}
\end{equation}

\section{Discussions}

Eq. (\ref{eq11}) provides an effective way of how to improve theoretical model through experimental results for nonlinear water wave theory. The parameter alpha in Eq. (\ref{eq11}) can be easily obtained from experimental tests. Another tool inherent in Eq. (\ref{eq11}) is that we may derive new equations mathematically for nonlinear wave equation by arbitrarily introducing higher order terms in Eq. (\ref{eq11}). This shows the power of the Replacement Rule. We will be explore this a little further here and will be detailed in subsequent papers.

The KdV equation gives waves travelling in the negative-direction for high wavenumbers (short wavelengths). This is in contrast with its purpose as an approximation for uni-directional waves propagating in the positive-direction. It has long been known that theoretical speed of a KdV solitary wave is not in perfect agreement with experimental results. In Fig. \ref{fig1} the experimental results from Daily and Stephan (1952), \cite{8} are reproduced with black dots to denote. At higher relative amplitude defined as the ratio of amplitude over water depth, phase speed deviates from the straight line defined by the dispersion relation Eq. (\ref{eq11}). Mei (1984) \cite{8} provided a more strict theory to improve theoretical results denoted in Fig. \ref{fig1} with dashed line. The best fit from the experimental tests is given 
\begin{equation}\tag{31}
c_p = c_0 \biggl [ 1+\frac{1}{2} \biggl( \frac{A}{h} \biggr) - \frac{3}{20}  \biggl( \frac{A}{h} \biggr)^2\biggr],
\end{equation}
from which and our Replacement Rule proposed herein, that is,
\begin{equation}\tag{32}
-i \omega A+i c_0 A k-i\frac{1}{6} c_0 h^2 A+ \frac{1}{2}i k c_0 A^2 -\frac{3}{20} i k c_0 A^3 =0,
\end{equation}
the corresponding governing equation should 
\begin{equation}\tag{32}
\eta_t +c_0 \eta_x +\frac{1}{6} c_0 h^2 A+ \frac{1}{2}i k c_0 A^2 -\frac{3}{20}i k c_0 A^3 =0,
\end{equation}
the corresponding governing equation should
\begin{equation}\tag{33}
\eta_t + c_0 \eta_x + \frac{1}{6}c_0 h^2 \eta_{xxx}+  \frac{1}{2h}c_0 \eta \eta_{x} - \frac{3}{20}  \frac{c_0}{h^2} \eta^{2} \eta_{x}=0.
\end{equation}
This is Gardner Equation obtained from experimental tests, which describes more accurate dispersion relation. We are thus led to the following interesting observations
\begin{enumerate}
\item 
KdV equation is important in the development of nonlinear water wave theory, but it may not be the right equation. The right equation should be Gardner equation, which permits soliton solution, undular bores, peakons etc. resembling true nonlinear water wave phenomena. 
\item
Gardner equation permits more solutions, thus able to describe more phenomena in nonlinear water waves. 
\end{enumerate}
In addition, these results add a deeper physical understanding of the meaning of dispersion relations for nonlinear waves. The integral in Eq. (4\ref{eq4}), representing the stationary-phase approximation, can be understood as the Cauchy’s residue theorem for complex integration around a semi-circle in the positive real-part semi-plane. This integral applied to water waves represents actually the expression of the Kramers-Kronig relations, \cite{14}, as the most general form of the dispersion relation. 

\section{Acknowledgement}

The present research is financially supported by the Natural Science Foundation of China under grants ($51639003 \& 51679037$).


\end{document}